\documentclass[letterpaper, inpress]{jds} % use this for production
\jdsmonth{October}                 % month
\jdsyear{2021}                  % year
\jdsvolume{xx}                  % volume number
\jdsissue{xx}                   % issue number
\jdsdoi{xx.xxxx/xxxxxxxxx}      % doi
\usepackage{amsfonts,amsmath,amssymb,amsthm}
\usepackage{booktabs}
\usepackage{longtable}

\usepackage{lipsum}

\title[Community Extraction of Text Networks]{Scalable Community Extraction of Text Networks for Automated Grouping in Medical Databases}

\author[1]{Tomilayo Komolafe}
\author[2]{Allan Fong}
\author[3]{Srijan Sengupta\footnote{Corresponding author. Email: ssengup2@ncsu.edu}}
\affil[1]{Qlik, 211 S Gulph Rd, King of Prussia, PA 19406}
\affil[2]{MedStar Health Research Institute, Hyattsville, Maryland 20782}
\affil[3]{Department of Statistics, North Carolina State University, Raleigh, NC, 27695}

\begin{document}

\maketitle

\begin{abstract}

Networks are ubiquitous in today's world.
{Community structure} is a well-known feature of many empirical networks, and a lot of statistical methods have been developed for community detection.
In this paper, we consider the problem of community structure in text networks, which is greatly relevant in medical errors and patient safety databases.
We adapt a well-known community extraction method to develop a scalable algorithm for community extraction in large text databases.
The application of our method on a real-world patient safety report database demonstrates that the groups generated from community extraction are much more accurate than manual tagging by frontline workers.
\end{abstract}


\section{Introduction}%
\label{sec:intro}
% Networks are capable of visually and mathematically representing a myriad of fields \citet{woodall2017overview} from the power grid (\citet{woodall2017overview}; \citet{albert2004structural}; \citet{dahan2017network}) where nodes are power stations and edges the transmission lines, to  a social network where nodes are individuals and interactions between individuals depicted as edges (\citet{dastour2016investigating}, or gene sequencing where the nucleotides that make up DNA and RNA during transcription are represented as network motifs (\citet{bacolla2011non}). We apply this novel approach to help tackle some of the challenges around patient safety and understanding medical errors.
Many complex systems in today's world consist, at an abstract level, of \textit{agents}
who \textit{interact} with one another.
% \citep{gavin2002functional,lynall2010functional}.
This general agent-interaction framework arises in a range of disciplines, such as 
biological sciences \citep{lynall2010functional},
physical sciences \citep{huberman1999internet,pagani2013power},  and social sciences \citep{milgram1967small}, to name a few.
By denoting \textit{agents} as \textit{nodes} and their \textit{interactions} as \textit{edges}, any such system can be represented as a network. 
Such network data provide a versatile framework for analyzing a broad spectrum of complex systems.

% Networks are ubiquitous in today's world, as a wide range of systems, such as  social interpersonal systems \citep{milgram1967small}, power grids \citep{watts1998collective}, and the World Wide Web \citep{huberman1999internet}, to name a few, can be represented as networks.
% %, and protein interaction systems \citep{gavin2002functional}, to name a few, can be represented as networks.
% Accordingly, there has been a lot of recent emphasis in the statistics literature towards developing statistical methodology for analyzing network data.
% %Broad overviews of network data analysis can be found in \cite{kolaczyk2009statistical}, \cite{goldenberg2010survey}, and \cite{newman2010networks}.
% Broad overviews of network data analysis can be found in \cite{kolaczyk2009statistical} and \cite{goldenberg2010survey}.

% \subsection{Community Structure in Networks}
{Community structure} is a well-known feature of many empirical networks.
Nodes in a network are often found to belong to groups or communities that
exhibit similar behavior.
The identification of this network structure, called {community detection}, is an important problem in network analysis.
Community detection has important scientific implications; these communities often turn out to be groups of agents which share common properties and/or play similar roles within the network.
For example, in \cite{jonsson2006cluster}, the communities in a protein interaction network turned out to be functional groups (proteins having the same or similar function) - this conclusion has important implications for cancer research.
\cite{fortunato2010community} provides a multidisciplinary exposition on community detection in networks.
Fittingly, several useful tools for community detection have been developed and studied in the statistics literature.
These include spectral methods \citep{rohe2011spectral, jin2015fast, sengupta2015spectral}, modularity based methods \citep{newman2004finding, bickel2009nonparametric,senguptapabm}, likelihood based methods \citep{amini2013}, to name a few.
Most of these methods are known to have theoretical guarantees for accuracy of community detection.

In this paper, we study text networks, where vertices represent documents and edges represent similarity between document pairs.
Similarity between text documents can be measured in a number of ways based on representational learning \citep{mikolov2013efficient,mikolov2017advances,hofmann1999probabilistic,landauer1998introduction,papadimitriou2000latent,dumais2004latent}.
We provide more details on document representation in Section \ref{sec:methods}.
Text networks provide a useful framework for representing large databases of documents, and statistical network analysis techniques can be applied for the analysis of such databases.
In particular, community detection techniques can be used for grouping text databases into homogeneous clusters, which enables downstream analysis of the clusters thus formed.
However,  there has not been much work on community detection of text networks, with some very recent exceptions such as \cite{yan2021overlapping} and \cite{dong2020overlapping}.

Our main contributions in this paper are as follows.
We develop a method for clustering text networks based on representational learning combined with a well-known community extraction method proposed by \cite{zhao2011community}.
 Most real-world text databases are large, which can lead to high computational expense when applying community extraction. We propose a novel divide and conquer strategy to address this issue.
   We demonstrate our method by applying it to a large patient safety event database, where it generates much better groups than manual tagging, as measured by document similarity.

The rest of the paper is structured as follows.
In Section \ref{sec:pse}, we describe the scientific application area of medical errors and patient safety events which motivated this work, and we also introduce the patient safety error database on which our method is applied.
In Section \ref{sec:methods}, we describe the proposed methodology.
In Section \ref{sec:results}, we report the results of our analysis, and we conclude the paper with a short discussion in Section \ref{sec:disc}.

% \ssg{TODO: a transition paragraph from networks and community structure to text networks and medical text databases. 
% Mention that there has not been much work on community detection in text networks.
% Some recent papers on this topic are \cite{yan2021overlapping,dong2020overlapping,de2016topic}.}

% \section{Text Networks in Medical Databases}

\section{Medical Errors and Patient Safety Event Reports}
\label{sec:pse}
The Institute of Medicine (IOM), an authority at the intersection of medicine and society, released a report titled “To Err is Human: Building a Safer Health System” in November 1999 \citep{donaldson2000err}. Its goal was to break the cycle of inaction regarding medical errors by advocating a comprehensive approach to improving patient safety. Based on two studies conducted in 1984 and 1992, the IOM concluded that between 44,000 and 98,000 patients die every year in United States (U.S.) hospitals due to medical errors. Costs alone from medical errors were approximated to be \$37.6 billion per year. About \$17 billion were associated with preventable errors \citep{donaldson2000err}. Given the intense level of public and scientific reaction to the report, various stakeholders responded swiftly to take action. In February 2000, President Clinton announced a national action plan to reduce preventable medical errors by fifty percent within five years \citep{Clinton2000}. Congress mandated the monitoring of progress in preventing patient harm. In July 2004, a Healthgrades Quality Study asserted that IOM had in fact vastly underestimated the number of deaths due to medical errors, citing 195,000 deaths per year \citep{harrington2005revisiting}.

Two decades later, medical errors continue to be a leading cause of death in the United States \citep{makary2016medical}. The Institute of Medicine and several state legislatures have recommended the use of patient safety event reporting systems (PSRS) to better understand and improve safety hazards (\citet{aspden2004patient}; \cite{rosenthal2005maximizing}). Numerous healthcare providers have adopted these systems which provide a framework for healthcare provider staff, including frontline clinicians, nurses, and technicians to report patient safety events, ranging from 'near misses', where no patient harm occurs, to serious safety events that result in patient harm \citep{clarke2006system}. However the potential of these reports to systematically identify hazards and reduce harm has been lacking, in part because of the limited techniques used to analyze these data. 
% Reported patient events range from \textit{near misses}, where no patient harm occurs, to serious safety events that result in patient harm. 
If the reported data can be analyzed effectively, reporting systems have the potential to dramatically improve the safety and quality of care by exposing possible weaknesses in the care process \citep{pronovost2008improving}.

% \subsection{Patient Safety Event Reports}
\textit{Patient safety event} (PSE) reports are free-text narratives written by the front-line staff. These narratives describe incidents whereby a healthcare service delivery did not go as expected. During these instances, the front-line staff witnessing the incident can document his/her perspective of the events that occurred. Therefore, aggregating similar PSEs has the potential to give insights into trends of the different types of medical errors healthcare organizations encounter. There is a significant amount of variation between documents because these narratives do not have to follow any specified format. For example, documents describing similar events can vary drastically in their word usage, vocabulary, document length, and prevalence of grammatical errors.
Therefore, the notion of similarity has to be based on semantic representation rather than simple features defined on the documents.

In this work, we consider a PSE database from MedStar Health consisting of 2,072 documents.
Our goal is to develop a clustering algorithm to find homogeneous groups of documents.
We now propose a method to accomplish this by using community extraction.

\section{Methodology}
\label{sec:methods}

\begin{figure}[htbp]
\centering
	\includegraphics[width=12cm,height=8cm]{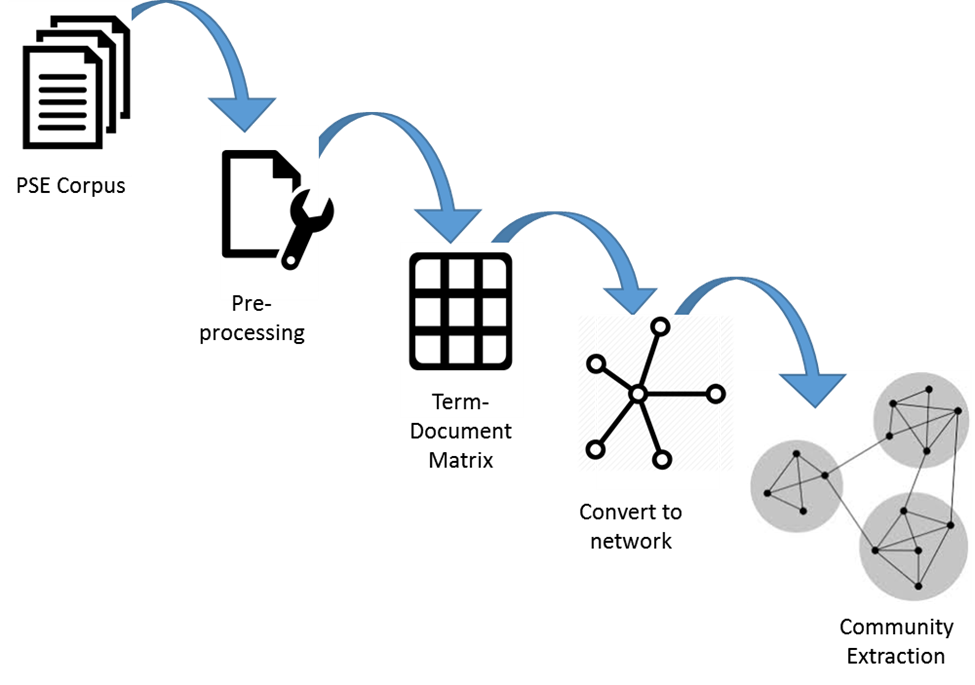}
 	\caption{Framework for community extraction of PSE corpus}
	 \label{framework}
\end{figure}

In this section, we describe the process of community extraction to find homogeneous clusters in a text database. 
Figure \ref{framework} provides a schematic representation of the different steps involved.
The subsequent subsections provide details on each step.
Note that while this work is motivated by patient safety event reports, this general methodology can be applied on any text database.
The first two steps (pre-processing and term-document matrix construction) are well-known strategies from natural language processing, while the last step (community extraction) is a well-known method from the statistical network analysis literature.
We integrate these well-known approaches in our work.
% Then we describe the community extraction framework and provide some recommendations to improve the method's performance for larger networks. 

\subsection{Text pre-processing}
A pre-processing step is critical to the performance of any natural language processing (NLP) model to reduce errors \citep{vijayarani2015preprocessing}. 
Pre-processing of text can be compared to exploratory data analysis in traditional statistical analysis.

In our work, we first create a manually curated dictionary of commonly misspelled words in our corpus and replace them with their proper spelling. To do this, we extracted terms that appeared in 2 or more documents and correct any misspelled terms.  As PSEs typically contain information such as the date an event occurred, the time it occurred, or dosage of a particular medication, any permutation of a date, dosage, or time is replaced with the words “date”, “dose”, and “time” respectively. This is because the exact time an event occurred or the exact dosage of a medication is irrelevant for our analysis.
However, we should not remove the word because then the sentence will lose its syntactic coherence.
Therefore, we simply replace specific times by the general concept word \textit{time}.
In addition, special characters are removed, except for periods and all other numbers are removed from the text.

For example, this sentence:
“\textit{On Dec. 13 at 5PM resident was prescribed 2mc/mg of oxycotine}” is converted to 
“\textit{On date at time resident was prescribed dose of oxycotine}”

 Furthermore, to ensure that words with similar morphology are presented as the same, we carried out stemming of the words, which is the process of reducing inflected (or sometimes derived) words to their word stem, base or root form—generally a written word form.
The goal of stemming is to reduce
inflectional forms and sometimes derivationally
related forms of a word to a common base form, and this is a common pre-processing step in text analytics \citep{vijayarani2015preprocessing}.

\subsection{Construction of Term Document Matrix}
The next step is to represent the text database as a numeric matrix.
This is a common approach in natural language processing, where the entire corpus is converted to a term-document matrix where rows represent terms and columns represent documents . The weighting of the terms in our term document matrix is critical to any future analysis. We use the common methodology Term Document - Inverse Document Frequency methodology referred to as “tf-idf” in the literature \citep{aizawa2003information,ramos2003using}. 
Here, \textit{term frequency} is an adjusted version of the number of times a term appears in the document.
Let $t$ be a term and $d$ be a document in the corpus.
Then, term frequency is defined as
\begin{equation}
    tf(t,d) = \frac{f_{t,d}}{\sum_{t' \in d} f_{t', d}},
\end{equation}
where $f_{t,d}$ is the raw count of a term in a document, i.e., the number of times that term t occurs in document d. 
Note that the denominator is the total frequency of all terms in the document, i.e., the total length of the document, which scales the raw count and allows for comparison between documents of differing length.
The inverse document frequency is a measure of how much information the word provides, i.e., how common or rare the term $t$ is across all documents in the corpus.
Let $D$ denote the set of all documents in the corpus and let $N = |D|$ be the total number of documents.
Then, inverse document frequency is defined as
\begin{equation}
idf(t,D) = \log\left(\frac{N}{|\{d \in D: t\in d \}|} \right),
\end{equation}
where the denominator is the number of documents which contain the term $t$.
Finally, the tf-idf score is calculated as
\begin{equation}
tf{\text -}idf(t,d) = tf(t,d) \times idf(t,D).
\end{equation}

The tf-idf term-document matrix is constructed as follows.
First, we consider the set of all unique terms that appear in the corpus.
Then, for each term $t$ and each document $d$, we compute the tf-idf score and populate the entries of the matrix.
For a toy illustration, consider the following short patient safety event reports.
\begin{itemize}
    \item "The patient schedule did not match the script."
    \item "Script and schedule mismatch. Script stated vasculab and schedule xray"
    \item "Xray monitor will not transmit images"
\end{itemize}
The resulting term-document matrix is displayed in Table \ref{tab:tfidf}.

\begin{table}[htbp]
  \caption{Toy illustration of TF-IDF matrix}%
  \label{tab:tfidf}
  \centering
  \begin{tabular}{lrrr}
    \toprule
    % \multicolumn{1}{l}{Season} & \multicolumn{1}{l}{Parameter}
    % & \multicolumn{2}{c}{Two-piece method} & \multicolumn{2}{c}{Marginal method} \\
    % \cmidrule(r){3-4} \cmidrule(r){5-6}
                              Terms & Doc1 & Doc2 & Doc3  \\
    \midrule
    did & 0.264 & 0.000 & 0.000 \\
    image & 0.000 & 0.000 & 0.264 \\
    match & 0.264 & 0.000 & 0.000 \\    
    mismatch & 0.000 & 0.176 & 0.000 \\
    monitor & 0.000 & 0.000 & 0.264 \\
    not & 0.097 & 0.000 & 0.097 \\
    patient & 0.097 & 0.065 & 0.000 \\    
    schedule & 0.097 & 0.130 & 0.000 \\
    script & 0.097 & 0.130 & 0.000 \\
    state & 0.000 & 0.176 & 0.000 \\    
    schedule & 0.000 & 0.000 & 0.000 \\
    transit & 0.000 & 0.000 & 0.264 \\
    vasculab & 0.000 & 0.176 & 0.000 \\    
    will & 0.000 & 0.000 & 0.264 \\
    xray & 0.000 & 0.065 & 0.097 \\

    \bottomrule
  \end{tabular}
\end{table}

\subsection{Text Network Construction via Latent Semantic Analysis}
Once we have a weighted term document matrix, we apply the well-known technique of Latent Semantic Analysis (LSA) for dimension reduction \citep{turney2001mining,dumais2004latent}. LSA has the ability to handle obstacles prevalent in natural language processing and analysis such as presence of synonyms and polysemy. In what follows, we provide  only a brief description of LSA. For a more detailed description of the approach, see \citet{landauer1998introduction}.

%  demonstrated that any rectangular matrix can be converted to a product of three matrices using the concept of singular value decomposition, or SVD as it is commonly referred to. \citep{golub1989cf}.
 For a term-document matrix $X$ of $m$ terms and $n$ documents with rank $r$, its singular value decomposition (SVD) can be written as 
\begin{equation}
  \label{eq:svd}
   X = T\Sigma D^T,
\end{equation}
where $X$ is the $m$ x $n$ term-document matrix, $T$ is a $m$ x $m$ matrix whose columns are the orthogonal eigenvectors of $XX^T$ where we denote $X^T$ as the transpose of the matrix $X$. The matrix $D$ is a $n$ x $n$ matrix whose columns are the orthogonal eigenvectors of $X^TX$ and $\Sigma$ is a $m$ x $n$ diagonal matrix whose diagonals are $\sqrt{\lambda_i}$ where $\lambda$ corresponds to the eigenvalues of $XX^T$ and 1 $\le$ $i$ $\le$ $r$ and 0 everywhere else.
The eigenvalues of $XX^T$ are the same as the eigenvalues of $X^TX$. The values $\sqrt{\lambda_i}$ are called the singular values of $X$. 

The implementation of LSA used in this work is a low rank approximation of the SVD. For this, we find a positive integer, $k$ $\le$ $r$ such that it closely approximates the term document matrix. The value $k$ is selected such that it minimizes the error between the original matrix $X$ and its low rank approximation $X_k$ . This is achieved through the following steps;
Since $\lambda_i$ $\ge$ $\lambda_{i+1}$, setting $\lambda_{i+1}$ $=$ $0$ if it is close to zero will not significantly affect the original matrix $X$. We therefore find a $k$ where 1 $\le$ $k$ $\le$ $r$ such that it minimizes the difference in the Frobenius norm between $X$ and $X_k$. If $k$ = $r$, then the difference in the Frobenius norm is 0 but if $k$ $\ll$ $r$, we have a low rank approximation of our matrix that is also easy to manipulate. By keeping only the $k$ columns or entries for each of our matrices, we obtain $X_k$ and furthermore a low rank approximation of both terms and documents. Therefore, we have 

\begin{equation}
  \label{eq:svdrnk}
   X_k = T_k\Sigma_k D_k^T
\end{equation}
Where we only keep the $k$ columns of matrix $T$ so $T_k$ is a $m$ x $k$ matrix, $D^T$ so $D_k^T$ is a $k$ x $n$ matrix and $\Sigma_k$ is a diagonal $k$ X $k$ matrix.
Then, the rows of the matrix $D_k$ are the LSA-based vector representations of the documents in the corpus.

Finally, we generate a network of documents by creating a similarity matrix from the matrix $D_k$. 
We define the similarity between two documents $d_i$ and $d_j$ as the correlation between the corresponding rows of $D_k$, resulting in a $n$ x $n$ correlation matrix. The correlation matrix serves as our adjacency matrix for the next step of community extraction. 
Note that this is a weighted adjacency matrix.

\subsection{Clustering of text network via community extraction}

Most community detection methods aim to partition a network into communities with the goal of maximizing the number of edges within communities and minimizing edges between communities. This framework assumes that all nodes belong to some community. However, there could be scenarios where some nodes do not belong to any particular community and forcing these nodes into a community will distort the community detection results. For example, let’s assume we have a network of high school students where links between students signifies that these students participate in similar extra-curricular activities. Applying some of the traditional community detection algorithms to this network will result in unsatisfactory results.  This is because some students naturally do not participate in any extra-curricular activity and therefore do not belong to a community. However, these community detection algorithms will force these nodes to one of the formed communities. 

The text networks from PSE databases also have this property.
We expect that the majority of PSE reports will fall into groups, but there could be some "miscellaneous" documents that do not belong to any group.
Community detection methods that partition all nodes into communities are going to enforce such "miscellaneous" reports into groups, which is unwarranted.
Therefore, we use the community extraction method, proposed by \citet{zhao2011community}, which can handle these types of networks. 

% Community extraction can handle these types of networks. \citet{zhao2011community} applied community extraction to some well-studied networks such as the karate club network \citet{zachary1977information},  \citet{hunter2008goodness} applied it to a school friendship network, and \citet{newman2006finding} applied community extraction to political books network . Their results show that community extraction performs better when compared to other popular community detection methods such as modularity using the approximate eigenvector solution \citet{newman2006finding}, \citet{newman2004finding}, fitting a block model via Markov chain Monte Carlo \citet{nowicki2001estimation}, or using the latent position cluster model \citet{handcock2007model}. In this paper, we use the community extraction method proposed by \citet{zhao2011community}. 

We describe a network graph $G$ as composed of vertices $V$ and edges $E$, and $G = (V, E)$. The total number of vertices in a network graph $G$ gives us the network size $N$. That is, $N$ $= |V|$. Also the number of edges in a network graph is $M$, where  $M = |E|$. We consider only non-overlapping communities in this paper, therefore once community extraction is applied to a network $G$ , the partition results in two distinct sets, $V_1$ and $V_2$ where $V_1$ $\cap$ $V_2$ = $\emptyset$ and $V_1$ $\cup$ $V_2$ = $V$.
A network can also be represented as an $N$ x $N$ adjacency matrix referred to as $\textbf{A}$, where its elements are $\textbf{A}_{ij}$ and $i,j = 1,2,...,N,$  $\textbf{A}_{ij}$ = $(-1,1)$ making it a weighted network. For text networks, the adjacency matrix $\textbf{A}$ is equal to the correlation matrix of $D_kT$ from the preceding subsection. Communities are extracted one at a time with the criterion of extracting a set of nodes with the sum of its weights largest within that set and smallest between the set and its complement \citep{zhao2011community}. We will call this set of extracted nodes $S$, and its complement, $S^c$. 
The objective function we are therefore maximizing in each iteration step is given by
\begin{equation}
  \label{eq:objfcn}
  \tilde{W}(S) = |S||S^c|\left[\frac{O(S)}{|S|^2} - \frac{B(S)}{|S||S^c|} \right],
\end{equation}
{where} $O(S) = \sum_{i,j \in S} A_{ij}$ and
  $B(S) = \sum_{i \in S,j \in S^c} A_{ij}$.
The term $O(S)$ is twice the weight of the edges within $S$ and $B(S)$ represents the weights from the set $S$ to the rest of the remaining network.  In large sparse networks, particularly as in our application, a small community $S$ could result in a large $\tilde{W}(S)$ value, the term $|S||S^c|$ serves to ensure that sufficiently sized communities are extracted at each step as very large communities or very small communities will be penalized. This is because the term, $|S||S^c|$ is maximized at $|S|$ = $\frac{N}{2}$.

% \begin{figure}[h!]
% \centering
% 	\includegraphics[width=15cm,height=10cm]{Figure 1 Framework.png}
%  	\caption{Framework for community extraction of PSE corpus applied in this work}
% 	 \label{framework}
% \end{figure}

%  Figure \ref{framework} describes the framework for community extraction applied to our corpus of PSE documents. 
 To maximize the objective function, we implement the \emph{tabu} search maximization technique which is a local optimization technique based on label switching \citep{beasley1998heuristic,glover1998tabu}.
In this optimization technique, a string of binary values representing nodes in either community $S$ or $S^c$  is passed to the \emph{tabu} search function  \citep{zhao2011community,beasley1998heuristic,glover1998tabu}. The function tracks which nodes have been switched, ensuring that they are not switched again until a certain number of iterations have passed, making these nodes, “\emph{tabu}”. To guard against being trapped at a local maxima, the algorithm is run with random label assignments each time.  

In our implementation, the community extraction algorithm is repeated till only a small subset of nodes, 30 nodes or less, are left in the network and this was sufficient for our application. \citet{zhao2011community} proposed a stopping criteria only for a network that can be represented by the block model. Future works will investigate a more appropriate stopping criteria.

\subsubsection{Scalability via Divide and Conquer Approach}

In practice, we observed run times of the order $O(n^2)$ where n  is the size of the corpus. Our original PSE corpus is 2,072 documents, and running one iteration of the \emph{tabu} search algorithm on the entire corpus takes over 120 hours.  One alternative is to use the divide and conquer strategy by splitting the entire corpus of 2,072 documents into chunks of 200 documents or chunks of 400 documents. We observed run times of about 22 hours and 44 hours when partitioned into sizes of 200 and 400 respectively. 

However, it is crucial to knit similar communities in each partition of 200 or 400 back together. 
Partitioning the entire document will also result in some communities being arbitrarily split up. We also developed a methodology for combining similar communities from different partitions. Our methodology relies on the correlation matrix of the entire 2,072 corpus. We compare pairs of communities across the different partitions and combine communities that have a combined density greater than some threshold. 

We denote $S_{a,p}$ as the identity of a community extracted during the implementation of our algorithm. The integer, $a$, refers to the iteration number at which the community is extracted in that partition. The integer, $p$, refers to the partition the community belongs to. Where 1 $\le$ $a$ $\le$ $x$ with $x$ representing the number of communities extracted for that partition and 1 $\le$ $p$ $\le$ $y$ where $y$ is the total number of partitions for that particular implementation. Therefore, to establish if two extracted communities, $S_{1,1}$ and  $S_{4,2}$ originally belonged to the same community, we compare each of their densities, $D_{a,p}$ to their combined density, $D_{(1,1),(4,2)}$. That is,
\begin{equation}
  D_{1,1} = \frac{1}{|S_{1,1}|^2}\sum_{i,j \in S_{1,1}} A_{ij},
  D_{4,2} = \frac{1}{|S_{4,2}|^2}\sum_{i,j \in S_{4,2}} A_{ij}, \text{ and }
  D_{(1,1),(4,2)} = \frac{1}{|S_{1,1}|*|S_{4,2}|}\sum_{i \in S_{1,1},j \in S_{4,2}} A_{ij}
\end{equation}
In this paper, two communities are combined together if $D_{(1,1)(4,2)}$ $>$ 0.85 * $D_{1,1}$ and $D_{(1,1)(4,2)}$ $>$ 0.85 * $D_{4,2}$.

\section{Empirical Results}%
\label{sec:results}
In this section, we report the results from applying the methodology proposed in Section \ref{sec:methods} on the MedStar PSE corpus of 2,072 documents.

\subsection{Benchmark Results from Manual Tagging}
First, we establish a reference method for benchmarking.
These PSE reports are manually tagged by the front-line staff with options available from a drop-down menu. Tags include both a general event description, and there are 20 options to select from in our report, and 187 specific event descriptions which are sub-categories of any one of the general event descriptions. If the tags are descriptive enough, then we would expect the diagonals of the correlation matrices, representing average correlation within a group, to be high, and conversely, the off diagonals to be low. This would suggest that front-line staff are tagging similar documents with similar tags. However, if the correlation matrices do not follow this pattern, then it suggests that the tags available to the front-line staff are not descriptive enough for each report type. 

\begin{figure}[htbp]
\centering
	\includegraphics[width=12cm,height=8cm]{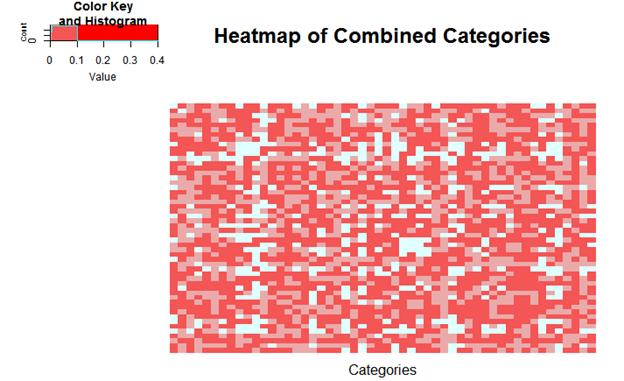}
 	\caption{Heatmap of groups generated by manual tagging from 52 X 52 combined event types, i.e., a General event tag paired with its Specific event.}
	 \label{corrCombEventTag}
\end{figure}

The benchmark results from manual tagging are displayed in Figure \ref{corrCombEventTag} as a heatmap.
% From our analysis, these manual tags do not properly describe the report. 
Clearly, manual tagging fails to obtain high correlation within groups and low correlation between groups.

Besides the visual illustrations, we can also look at statistics of the correlation matrices obtained by manual tagging.
Specifically, are there communities or tags whereby the documents within the community are more related to another set of documents in another community or tag. We do this by looking at the percentage of off diagonal cells that have a value equal to or greater than the value of the cell in the diagonal for a given column in the correlation matrix. 
Some examples of manually tagged categories that are more similar to other categories than within themselves are below. 

\begin{itemize}
    \item ``Medication'': more related with ``Fluid-Outdated'' and ``Unusable Medication''
    \item ``Equipment'': more related with ``Medical Device-Sterilization'' and ``Cleanliness Issue''
    \item ``Diagnostic Imaging-Test - Wrong Side (L vs. R)'': more related with ``Blood Bank-Patient Testing (Blood Bank Use Only)'', ``Diagnostic Imaging-Image - Misidentified'', and ``Diagnostic Imaging-Test - Test Delayed''
\end{itemize}

\subsection{Results from Community Extraction}
Next, we applied our methodology described in Section \ref{sec:methods} to obtain automated tags via community extraction.
Recall that implementing the method on the full network of 2,072 documents is computationally very intensive, and therefore we applied the divide and conquer approach described at the end of Section \ref{sec:methods}.
We used subnetworks of $200$ or $400$ documents, and used a correlation cut-off of $0.15$ or $0.2$ to stop the network from getting too dense.

\begin{figure}[htbp]
\centering
	\includegraphics[width=10cm,height=6cm]{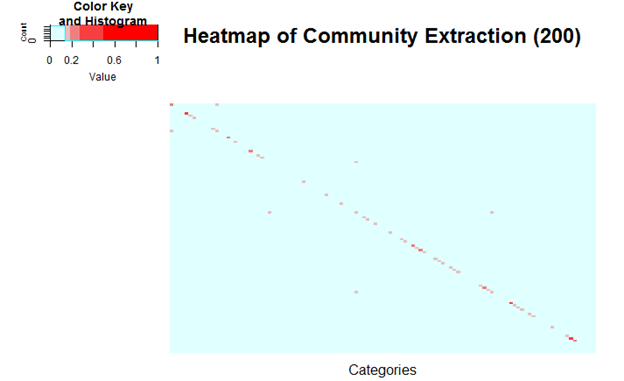}
	\includegraphics[width=10cm,height=6cm]{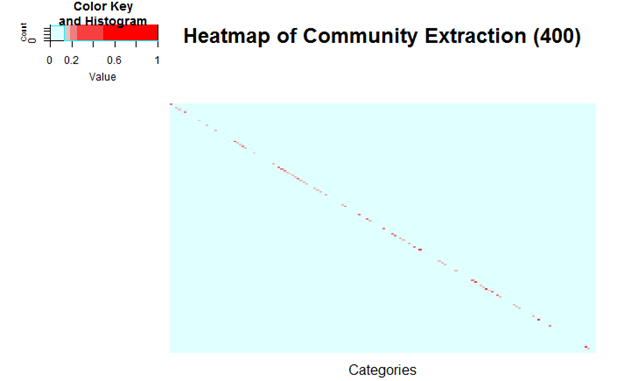}
	\includegraphics[width=10cm,height=6cm]{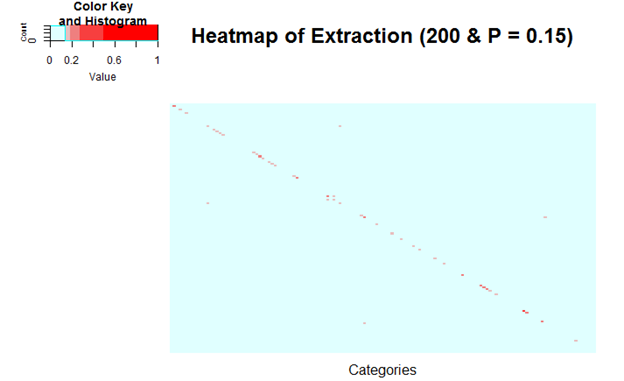}
 	\caption{Heatmap of communities generated from correlation matrix of documents that fall into the respective communities after community extraction is applied. 
 	Top: Partitions of 200 documents with threshold 0.2 and 113 communities;
 	Middle: Partitions of 400 documents with threshold 0.2 and 156 communities;
 	Bottom: Partitions of 200 documents with threshold 0.15 and 125 communities.}
	 \label{corrExt_200_02}
\end{figure}

The results are plotted in Figure \ref{corrExt_200_02}.
Note that the community extraction method does not require pre-specification of the number of communities, rather, the number of communities is an output of the method.
We obtained $113$, $156$, and $125$ clusters, respectively, from top to bottom of Figure \ref{corrExt_200_02}.
From the correlation heatmaps, it is clear that the documents have very high within-group correlation and very low between-group correlation, which indicates that the grouping is effective.
This is a substantial improvement over manual tagging (Figure \ref{corrCombEventTag}).
Note that the results from community extraction are better than manual tagging across the range of tuning parameters, i.e., subgraph size and correlation threshold.

\begin{figure}[htbp]
\centering
	\includegraphics[width=12cm,height=8cm]{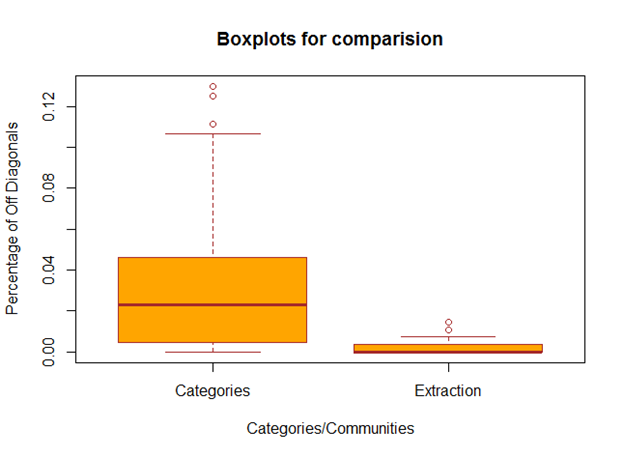}
 	\caption{Comparisons of communities between the predefined PSE categories/tag against community extraction by looking at the distribution of percentage of communities that are more similar, higher correlation score, than documents within that community.}
	 \label{Boxplots}
\end{figure}

Next, recall that we observed ``heterophilic'' behavior with manual tagging, where documents in some groups have higher between-group correlation than within-group correlation.
To compare manual tagging vs community extraction with respect to this property, we looked at each group, and computed what fraction of other groups have higher between-group correlation than within-group correlation.
The boxplots are shown in Figure \ref{Boxplots}, where we compare manual tagging to a representative community extraction.
We observe that the groups from community extraction have very little ``heterophilic'' behavior comapred to manual tagging.

% \begin{figure}[h!]
% \centering
% 	\includegraphics[width=15cm,height=10cm]{Figure 4 Corr Matrix 2.png}
%  	\caption{Heatmap of groups generated from automated tagging via community extraction and divide and conquer. 
%  	Top: Divide anmd156  X 156 correlation matrix of documents that fall into the respective communities after community extraction is applied. Partition (400) with correlation threshold set at 0.2}
% 	 \label{corrExt_400_02}
% \end{figure}

% \begin{figure}[h!]
% \centering
% 	\includegraphics[width=15cm,height=10cm]{Figure 5 Corr Matrix 2.png}
%  	\caption{Heatmap of communities generated from 125  X 125 correlation matrix of documents that fall into the respective communities after community extraction is applied. Partition (400) with correlation threshold set at 0.15}
% 	 \label{corrExt_200_015}
% \end{figure}

\begin{figure}[htbp]
\centering
	\includegraphics[width=12cm,height=8cm]{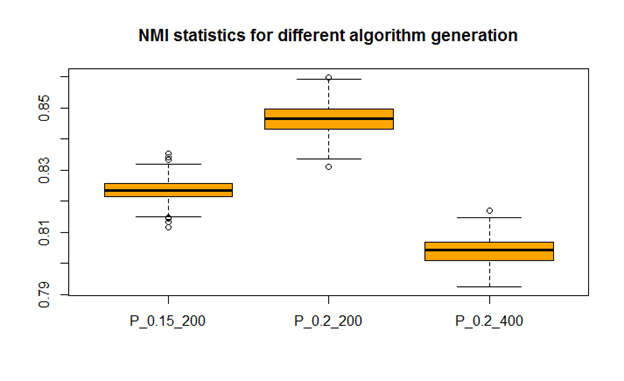}
 	\caption{Normal Mutual Information (NMI) statistics for comparing the relatedness of communities extracted for the different permutations of partition size and correlation matrix threshold.}
	 \label{NMIExt_Thresh_Part}
\end{figure}

Finally, recall that our divide and conquer strategy involves random partitioning of the large text network into a number of smaller subnetworks.
A natural question is "How stable are the groupings generated due to random partitioning? To answer this question, we implemented several random iterations of the divide and conquer strategy, and computed the Normalized Mutual Information (NMI) for document groups arising in different iterations. A high value of NMI indicates high stability of document grouping across random iterations.
The results are plotted in Figure \ref{NMIExt_Thresh_Part} for several tuning parameter values.
We observe that the NMI values are quite high indicating stability of clustering.

% \begin{figure}[h!]
% \centering
% 	\includegraphics[width=10cm,height=6cm]{Figure 9 NMI Histogram Distribution.png}
%  	\caption{Distribution of NMI results for partition size of 400 and correlation threshold of 0.2}
% 	 \label{HistNMIPart400}
% \end{figure}

\section{Discussion}
\label{sec:disc}

\subsection{Emerging communities and themes}
This analysis demonstrates the advantage of a network driven approach to extract communities in patient safety event free-text. The results clearly show categories with less overlapping categories compared to the categories manually selected by the front-line staff. It is likely that the analysis can identify communities of reports or themes in the reports that are not dependent on the structured categories. For example, communication and hand-off are often prevalent themes in patient safety reports that are not typically captured in structured fields. 
Structured fields are predefined and often difficult or time consuming to change and update. A network driven approach that leverages the free-text is more flexible and can identify more timely hazards with changing environments and care processes. Having a flexible approach is particularly important as new workflows are being introduced (e.g., COVID-19 protocols, telehealth).

\subsection{Opportunities to improve reporting and analysis}
A network driven approach to identify communities and themes in free-text can help reduce the burden of reporters from choosing through complex taxonomies which are both time consuming and can result in errors. In addition, these results highlight the potential to identify communities of related reports that might be missed from analyzing just the structured categories. Such categorization flexibility could greatly help safety analysts and safety leaders better identify meaningful signals and insights from all the data.

\subsection{Limitations}
This analysis was performed on data from one healthcare system. As a result, the comparison of extracted communities with the structured categories are specific to the structured categories implemented at the healthcare system. It is possible that other healthcare systems use different categorization taxonomies highlighting the need to understand the generalizability of this approach across taxonomies and healthcare systems. In addition, the present method does not consider temporal effects on communities. Expanding this approach to include temporally stable communities or emerging communities would be important especially as changes to policy, workflow, safety hazards can often occur.

\bibliographystyle{jds}
\bibliography{JDSbib,ref}

\begin{thebibliography}{36}
\providecommand{\natexlab}[1]{#1}

\bibitem[{Aizawa(2003)}]{aizawa2003information}
Aizawa A (2003).
\newblock An information-theoretic perspective of tf--idf measures.
\newblock \emph{Information Processing \& Management}, 39(1): 45--65.

\bibitem[{Amini et~al.(2013)Amini, Chen, Bickel, and Levina}]{amini2013}
Amini AA, Chen A, Bickel PJ, Levina E (2013).
\newblock Pseudo-likelihood methods for community detection in large sparse
  networks.
\newblock \emph{Ann. Statist.}, 41(4): 2097--2122.

\bibitem[{Aspden et~al.(2004)Aspden, Corrigan, Wolcott, Erickson
  et~al.}]{aspden2004patient}
Aspden P, Corrigan JM, Wolcott J, Erickson SM, et~al. (2004).
\newblock Patient safety reporting systems and applications.
\newblock In: \emph{Patient Safety: Achieving a New Standard for Care}.
  National Academies Press (US).

\bibitem[{Beasley(1998)}]{beasley1998heuristic}
Beasley JE (1998).
\newblock Heuristic algorithms for the unconstrained binary quadratic
  programming problem.
\newblock \emph{Technical report}, Citeseer.

\bibitem[{Bickel and Chen(2009)}]{bickel2009nonparametric}
Bickel PJ, Chen A (2009).
\newblock A nonparametric view of network models and {N}ewman--{G}irvan and
  other modularities.
\newblock \emph{Proceedings of the National Academy of Sciences}, 106:
  21068--21073.

\bibitem[{Clarke(2006)}]{clarke2006system}
Clarke JR (2006).
\newblock How a system for reporting medical errors can and cannot improve
  patient safety.
\newblock \emph{The American Surgeon}, 72(11): 1088--1091.

\bibitem[{Donaldson et~al.(2000)Donaldson, Corrigan, Kohn
  et~al.}]{donaldson2000err}
Donaldson MS, Corrigan JM, Kohn LT, et~al. (2000).
\newblock To err is human: building a safer health system.

\bibitem[{Dong et~al.(2020)Dong, Yang, and Chen}]{dong2020overlapping}
Dong R, Yang J, Chen Y (2020).
\newblock Overlapping community detection in weighted temporal text networks.
\newblock \emph{IEEE Access}, 8: 58118--58129.

\bibitem[{Dumais(2004)}]{dumais2004latent}
Dumais ST (2004).
\newblock Latent semantic analysis.
\newblock \emph{Annual review of information science and technology}, 38(1):
  188--230.

\bibitem[{Fortunato(2010)}]{fortunato2010community}
Fortunato S (2010).
\newblock Community detection in graphs.
\newblock \emph{Physics Reports}, 486(3): 75--174.

\bibitem[{Glover and Laguna(1998)}]{glover1998tabu}
Glover F, Laguna M (1998).
\newblock Tabu search.
\newblock In: \emph{Handbook of combinatorial optimization}, 2093--2229.
  Springer.

\bibitem[{Harrington(2005)}]{harrington2005revisiting}
Harrington MM (2005).
\newblock Revisiting medical error: Five years after the iom report, have
  reporting systems made a measurable difference.
\newblock \emph{Health Matrix}, 15: 329.

\bibitem[{Hofmann(1999)}]{hofmann1999probabilistic}
Hofmann T (1999).
\newblock Probabilistic latent semantic indexing.
\newblock In: \emph{Proceedings of the 22nd annual international ACM SIGIR
  conference on Research and development in information retrieval}, 50--57.

\bibitem[{HOUSE(2020)}]{Clinton2000}
HOUSE TW (2020).
\newblock Clinton-gore administration announces new actions to improve patient
  safety and assure health care quality.
\newblock
  \url{https://clintonwhitehouse4.archives.gov/textonly/WH/New/html/20000222_1.html}.

\bibitem[{Huberman and Adamic(1999)}]{huberman1999internet}
Huberman BA, Adamic LA (1999).
\newblock Internet: growth dynamics of the {W}orld-{W}ide {W}eb.
\newblock \emph{Nature}, 401: 131.

\bibitem[{Jin(2015)}]{jin2015fast}
Jin J (2015).
\newblock Fast community detection by {SCORE}.
\newblock \emph{The Annals of Statistics}, 43(1): 57--89.

\bibitem[{Jonsson et~al.(2006)Jonsson, Cavanna, Zicha, and
  Bates}]{jonsson2006cluster}
Jonsson PF, Cavanna T, Zicha D, Bates PA (2006).
\newblock Cluster analysis of networks generated through homology: automatic
  identification of important protein communities involved in cancer
  metastasis.
\newblock \emph{BMC Bioinformatics}, 7(1): 2.

\bibitem[{Landauer et~al.(1998)Landauer, Foltz, and
  Laham}]{landauer1998introduction}
Landauer TK, Foltz PW, Laham D (1998).
\newblock An introduction to latent semantic analysis.
\newblock \emph{Discourse processes}, 25(2-3): 259--284.

\bibitem[{Lynall et~al.(2010)Lynall, Bassett, Kerwin, McKenna, Kitzbichler,
  Muller et~al.}]{lynall2010functional}
Lynall ME, Bassett DS, Kerwin R, McKenna PJ, Kitzbichler M, Muller U, et~al.
  (2010).
\newblock Functional connectivity and brain networks in schizophrenia.
\newblock \emph{Journal of Neuroscience}, 30(28): 9477--9487.

\bibitem[{Makary and Daniel(2016)}]{makary2016medical}
Makary MA, Daniel M (2016).
\newblock Medical error—the third leading cause of death in the us.
\newblock \emph{Bmj}, 353.

\bibitem[{Mikolov et~al.(2013)Mikolov, Chen, Corrado, and
  Dean}]{mikolov2013efficient}
Mikolov T, Chen K, Corrado G, Dean J (2013).
\newblock Efficient estimation of word representations in vector space.
\newblock \emph{arXiv preprint arXiv:1301.3781}.

\bibitem[{Mikolov et~al.(2017)Mikolov, Grave, Bojanowski, Puhrsch, and
  Joulin}]{mikolov2017advances}
Mikolov T, Grave E, Bojanowski P, Puhrsch C, Joulin A (2017).
\newblock Advances in pre-training distributed word representations.
\newblock \emph{arXiv preprint arXiv:1712.09405}.

\bibitem[{Milgram(1967)}]{milgram1967small}
Milgram S (1967).
\newblock The small world problem.
\newblock \emph{Psychology Today}, 2: 60--67.

\bibitem[{Newman and Girvan(2004)}]{newman2004finding}
Newman MEJ, Girvan M (2004).
\newblock Finding and evaluating community structure in networks.
\newblock \emph{Physical review E}, 69(2): 026113.

\bibitem[{Pagani and Aiello(2013)}]{pagani2013power}
Pagani GA, Aiello M (2013).
\newblock The power grid as a complex network: a survey.
\newblock \emph{Physica A: Statistical Mechanics and its Applications},
  392(11): 2688--2700.

\bibitem[{Papadimitriou et~al.(2000)Papadimitriou, Raghavan, Tamaki, and
  Vempala}]{papadimitriou2000latent}
Papadimitriou CH, Raghavan P, Tamaki H, Vempala S (2000).
\newblock Latent semantic indexing: A probabilistic analysis.
\newblock \emph{Journal of Computer and System Sciences}, 61(2): 217--235.

\bibitem[{Pronovost et~al.(2008)Pronovost, Morlock, Sexton, Miller,
  Holzmueller, Thompson et~al.}]{pronovost2008improving}
Pronovost PJ, Morlock LL, Sexton JB, Miller MR, Holzmueller CG, Thompson DA,
  et~al. (2008).
\newblock Improving the value of patient safety reporting systems.
\newblock \emph{Advances in Patient Safety: New Directions and Alternative
  Approaches (Vol. 1: Assessment)}.

\bibitem[{Ramos et~al.(2003)}]{ramos2003using}
Ramos J, et~al. (2003).
\newblock Using tf-idf to determine word relevance in document queries.
\newblock In: \emph{Proceedings of the first instructional conference on
  machine learning}, volume 242, 29--48. Citeseer.

\bibitem[{Rohe et~al.(2011)Rohe, Chatterjee, and Yu}]{rohe2011spectral}
Rohe K, Chatterjee S, Yu B (2011).
\newblock Spectral clustering and the high-dimensional stochastic blockmodel.
\newblock \emph{The Annals of Statistics}, 39(4): 1878--1915.

\bibitem[{Rosenthal and Booth(2005)}]{rosenthal2005maximizing}
Rosenthal J, Booth M (2005).
\newblock \emph{Maximizing the use of state adverse event data to improve
  patient safety}.
\newblock National Academy for State Health Policy Portland, ME.

\bibitem[{Sengupta and Chen(2015)}]{sengupta2015spectral}
Sengupta S, Chen Y (2015).
\newblock Spectral clustering in heterogeneous networks.
\newblock \emph{Statistica Sinica}, 25: 1081--1106.

\bibitem[{Sengupta and Chen(2018)}]{senguptapabm}
Sengupta S, Chen Y (2018).
\newblock A block model for node popularity in networks with community
  structure.
\newblock \emph{Journal of the Royal Statistical Society: Series B (Statistical
  Methodology)}, 80(2): 365--386.

\bibitem[{Turney(2001)}]{turney2001mining}
Turney PD (2001).
\newblock Mining the web for synonyms: Pmi-ir versus lsa on toefl.
\newblock In: \emph{European conference on machine learning}, 491--502.
  Springer.

\bibitem[{Vijayarani et~al.(2015)Vijayarani, Ilamathi, Nithya
  et~al.}]{vijayarani2015preprocessing}
Vijayarani S, Ilamathi MJ, Nithya M, et~al. (2015).
\newblock Preprocessing techniques for text mining-an overview.
\newblock \emph{International Journal of Computer Science \& Communication
  Networks}, 5(1): 7--16.

\bibitem[{Yan et~al.(2021)Yan, Jia, and Wang}]{yan2021overlapping}
Yan S, Jia Y, Wang X (2021).
\newblock Overlapping community detection in temporal text networks.

\bibitem[{Zhao et~al.(2011)Zhao, Levina, and Zhu}]{zhao2011community}
Zhao Y, Levina E, Zhu J (2011).
\newblock Community extraction for social networks.
\newblock \emph{Proceedings of the National Academy of Sciences}, 108(18):
  7321--7326.

\end{thebibliography}

\section{Appendix}

\begin{center}
\begin{longtable}{|c l|}
\caption{List of General Event Tags from PSE} \label{tab:SpecifEvnt1} \\

\hline \multicolumn{1}{|c}{\textbf{}} &  \multicolumn{1}{l|}{\textbf{ General Event Types }} \\ \hline 
\endfirsthead

\multicolumn{2}{l}%
{{\bfseries \tablename\ \thetable{} -- continued from previous page}} \\
\hline \multicolumn{1}{|c}{\textbf{}} &  \multicolumn{1}{l|}{\textbf{ General Event Types }} \\ \hline 
\endhead

\hline \multicolumn{2}{|r|}{{Continued on next page}} \\ \hline
\endfoot

\hline \hline
\endlastfoot
    1 & Airway Management  \\
    2 & Blood Bank \\
    3 & Diagnosis/Treatment  \\    
    4 & Diagnostic Imaging\\
    5 & Equipment/Medical Device\\
    6 & Facilities\\
    7 & Fall\\
    8 & Healthcare IT\\
    9 &  Infection Prevention\\
    10 &  Lab/Specimen\\
    11 &  Lines/Tubes/Drain\\
    12 &  Maternal/Childbirth\\
    13 &  Medication/Fluid\\
    14 &  Miscellaneous\\
    15 &  Patient ID/Documentation/Consent\\
    16 &  Professional Conduct\\
    17 &  Restraints/Seclusion Injury\\
    18 &  Safety/Security\\
    19 &  Skin/Tissue\\
    20 &  Surgery/Procedure\\
    \end{longtable}
\end{center}

\clearpage
\begin{center}
\begin{longtable}{|c l|}
\caption{List of Specific Event Types from PSE} \label{tab:SpecifEvnt2} \\

\hline \multicolumn{1}{|c}{\textbf{}} &  \multicolumn{1}{l|}{\textbf{Specific Event Types}} \\ \hline 
\endfirsthead

\multicolumn{2}{l}%
{{\bfseries \tablename\ \thetable{} -- continued from previous page}} \\
\hline \multicolumn{1}{|c}{\textbf{}} &  \multicolumn{1}{l|}{\textbf{Specific Event Types}} \\ \hline 
\endhead

\hline \multicolumn{2}{|r|}{{Continued on next page}} \\ \hline
\endfoot

\hline \hline
\endlastfoot
    1 & Abandonment  \\
    2 & Abrasion \\
    3 & Abuse/Assault (Physical)  \\    
    4 & Abuse/Assault (Verbal)  \\    
    5 & Administration Technique Incorrect  \\    
    6 & Adverse Drug Reaction  \\    
    7 & Adverse Reaction (Non Med)  \\    
    8 & Air Quality/Odor/Smoke/Fumes  \\    
    9 & Airway Mgmt Equipment Issue  \\    
    10 & Airway Obstructed  \\    
    11 & Apgar Score < 5 at 5 min  \\    
    12 & Armband Issue  \\    
    13 & Bed Malfunction  \\    
    14 & Birth Trauma  \\    
    15 & Blister  \\    
    16 & Break in Sterile Technique  \\    
    17 & Broken Item  \\    
    18 & Bruise  \\    
    19 & Burn  \\    
    20 & Cardiac and/or Respiratory Arrest Requiring ACLS Intervention  \\    
    21 & Cardiac or Circulatory Event  \\    
    22 & Cardiopulmonary Arrest Outside of ICU Setting  \\    
    23 & Circulation Impeded  \\    
    24 & Collection Issue  \\    
    25 & Combination or Interaction of Device Defect and Use Error \\    
    26 & Communication  \\    
    27 & Complications of Anesthesia  \\    
    28 & Complications of Surgery/Procedure  \\    
    29 & Consent Issue  \\    
    30 & Contamination  \\    
    31 & Contrast/Radiopharmaceutical - Allergic Reaction  \\    
    32 & Contrast/Radiopharmaceutical - Event  \\  
    33 & Contrast/Radiopharmaceutical - Extravisation  \\  
    34 & Count Issue  \\  
    35 &
    Date of Birth Issue \\  
    36 &
    Delay/Difficulty With Resuscitation \\  
    37 &
    Delivery Without Provider \\  
    38 &
    Diagnosis - Delayed \\  
    39 &
    Diagnosis - Missed \\  
    40 &
    Diagnosis Issue \\  
    41 &
    Diaper Dermatitis \\  
    42 &
    Dietary Issue \\  
    43 &
    Disconnected \\  
    44 &
    Discontinued \\  
    45 &
    Discontinued Incorrectly \\  
    46 &
    Dislodgement \\  
    47 &
    Disorderly Person \\  
    48 &
    Disrupted Utility (Electric/Water/HVAC/Med Gas) \\  
    49 &
    Documentation Error \\  
    50 &
    Documentation Issue \\  
    51 &
    Dose/Concentration Incorrect \\  
    52 &
Drug Incorrect \\ 
53 &
Drug Interaction/Incompatibility \\ 
54 &
Drug Preparation/Labeling Issue \\ 
55 &
Drug With Known Allergy \\ 
56 &
Duplicate Therapy \\ 
57 &
Elevator Malfunction \\ 
58 &
Elopement \\ 
59 &
Equipment - Faulty \\ 
60 &
Equipment - Not Available \\ 
61 &
Equipment - Wrong/Inappropriate \\ 
62 &
Equipment (Blood Bank Use Only) \\ 
63 &
Equipment/Device Function \\ 
64 &
Exposure - Prolonged Fluro Time \\ 
65 &
Extubation - Unplanned \\ 
66 &
Extubation Issue - Self \\ 
67 &
Failure to Assess Patient \\ 
68 &
Failure to Follow Order \\ 
69 &
Failure to Respond to Request for Service \\ 
70 &
Fetal pH <7.05 Cord Blood Gas \\ 
71 &
Foreign Object Retained Post Procedure \\ 
72 &
Friction/Shear \\ 
73 &
From Bed \\ 
74 &
From Bed - Over Rails \\ 
75 &
From Chair \\ 
76 &
From Exam Stool \\ 
77 &
From Exam/Operating Table \\ 
78 &
From Stretcher \\ 
79 &
From Therapy Equipment \\ 
80 &
From Toilet/Commode \\ 
81 &
From Wheelchair \\ 
82 &
Hand Hygiene Compliance Issue \\ 
83 &
Hardware Failure or Problem \\ 
84 &
Illegible Order \\ 
85 &
Image - Misidentified \\ 
86 &
Implant Issue \\ 
87 &
Inadequate Supplies \\ 
88 &
Inappropriate Admission \\ 
89 &
Inappropriate Discharge \\ 
90 &
Inconsiderate/Rude/Hostile/Inappropriate Behaviors \\ 
91 &
Infiltration Event \\ 
92 &
Infiltration/Extravasation \\ 
93 &
Intimidation/Verbal Abuse \\ 
94 &
Intubation - Unplanned \\ 

95 &
Isolation - Failure to Follow Protocol \\ 
96 &
Labeling Issue \\ 
97 &
Laceration \\ 
98 &
Lack of Responsiveness \\ 
99 &
Left Against Medical Advice \\ 
100 &
Left Without Being Seen \\ 
101 &
Line Not Changed \\ 
102 &
Lost Specimen \\ 
103 &
Medication Administered Not Ordered \\ 
104 &
Monitoring Issue \\ 
105 &
MRI Safety Issue \\ 
106 &
Narcotic Count Incorrect \\ 
107 &
Network Failure or Problem \\ 
108 &
Non Head Injury - Restraint Related \\ 
109 &
Noncompliant/Uncooperative/Obstructive Behaviors \\ 
110 &
Not Activating the Chain of Command \\ 
111 &
Occlusion \\ 
112 &
Omission \\ 
113 &
Ordering Issue \\ 
114 &
Other (please specify) \\ 
115 &
Outdated/Unusable Medication \\ 
116 &
Patient Exposure - Blood/Body Fluid \\ 
117 &
Patient Testing (Blood Bank Use Only) \\ 
118 &
Personal/Associate Property Lost/Theft \\ 
119 &
Phlebitis \\ 
120 &
Post-Partum Hemorrhage \\ 
121 &
Preparation Incorrect \\ 
122 &
Prescriptions Not Given at Discharge \\ 
123 &
Pressure Ulcer \\ 
124 &
Procedure Issue \\ 
125 &
Process Issue \\ 
126 &
Product Administration (Clinical Services) \\ 
127 &
Product Receipt/Handling (Blood Bank Use Only) \\ 
128 &
Product Test Request (Clinical Services) \\ 
129 &
Property Damage/Vandalism \\ 
130 &
Pump Programming Issue \\ 
131 &
Radiation Onclogy Issues \\ 
132 &
Referral Issue \\ 
133 &
Reporting Issue \\ 
134 &
Requisition Incorrect \\ 
135 &
Respiratory Mgmt - Inappropriate \\ 
136 &
Restraint Improperly Applied \\ 
137 &
Restraints Applied - Not Ordered \\ 
138 &
Restraints Ordered - Not Applied \\ 
139 &
Results - Delay in Critical Results Communication \\ 
140 &
Results - Posted to Wrong Patient \\ 
141 &
Risky/Reckless/Dangerous Behaviors \\ 
142 &
Route Incorrect \\ 
143 &
Sample \\ 
144 &
Self Injury \\ 
145 &
Shoulder Dystocia \\ 
146 &
Site Infection \\ 
147 &
Skin Tear \\ 
148 &
Slip/Trip/Fall \\ 
149 &
Smoking \\ 
150 &
Specimen Acceptability Issue \\ 
151 &
Specimen Processing Issue \\ 
152 &
Sterilization/Cleanliness Issue \\ 
153 &
Storage Incorrect \\ 
154 &
Suicide/Suicide AttemptSuspicious Package \\ 
155 &
Test - Incorrectly Performed \\ 
156 &
Test - Ordered, Not Performed \\ 
157 &
Test - Test Delayed \\ 
158 &
Test - Wrong Side (L vs. R) \\ 
159 &
Testing Issue \\ 
160 &
Time/Date Incorrect/Delayed \\ 
161 &
Tissue \\ 
162 &
Treatment - Delayed \\ 
163 &
Treatment - Inappropriate \\ 
164 &
Treatment - Incorrectly Performed \\ 
165 &
Treatment - No Order for \\ 
166 &
Unable to Access \\ 
167 &
Unauthorized Access/Trespassing \\ 
168 &
Unauthorized Drugs \\ 
169 &
Unauthorized Weapons on Premises \\ 
170 &
Unexpected Return to the OR \\ 
171 &
Unexpected Software Design Issue \\ 
172 &
Unexpected Transfer to ICU/NICU \\ 
173 &
Unknown/Found on Floor \\ 
174 &
Use Error \\ 
175 &
Visitor Policy Issue \\ 
176 &
Water Leak/Flood \\ 
177 &
Weapons on Premises \\ 
178 &
While Ambulating \\ 
179 &
While Held by Staff \\ 
180 &
While Running/Playing \\ 
181 &
While Standing \\ 
182 &
While Transferring \\ 
183 &
Workplace Violence \\ 
184 &
Wound \\ 
185 &
Wrong Body Part (Site/Side/Level) \\ 
186 &
Wrong Insertion Location \\ 
187 &
Wrong Patient \\ 
\end{longtable}
\end{center}
\end{document}